\documentclass[aps,prb,preprint,showpacs]{revtex4}

\usepackage{amsmath,bm,amsfonts}
\usepackage{graphicx}

\begin{document}

\title{Dual Spin Filter Effect in a Zigzag Graphene Nanoribbon}

\author{Taisuke Ozaki$^{1}$, Kengo Nishio$^{2}$, Hongming Weng$^{1}$, and Hiori Kino$^{3}$}
\address{
     $^{1}$Research Center for Integrated Science (RCIS), Japan 
     Advanced Institute of Science and Technology (JAIST), 
     1-1 Asahidai, Nomi, Ishikawa 923-1292 Japan,\\
     $^{2}$Research Institute for Computational Sciences (RICS),
     National Institute of Advanced
     Industrial Science and
     Technology (AIST),
     1-1-1 Umezono, Tsukuba,
     Ibaraki 305-8568, Japan,\\
     $^{3}$National Institute for Material Science (NIMS),
     1-2-1 Sengen, Tsukuba,
     Ibaraki 305-0047, Japan
}

\date{\today}

\begin{abstract} 
 By first principle calculations, a dual spin filter effect 
 under finite bias voltages is demonstrated in an antiferromagnetic 
 junction of symmetric zigzag graphene nanoribbon (ZGNR). 
 Unlike conventional spin filter devices using half metallic materials,
 the up- and down-spin electrons are unidirectionally filtered in 
 the counter direction of the bias voltage, making the junction 
 a dual spin filter. On the contrary, asymmetric ZGNRs do not exhibit
 such a spin filter effect.
 By analyzing Wannier functions and a tight-binding model,
 we clarify that 
 an interplay between the spin polarized band structure of $\pi$ 
 and $\pi^*$ states near the Fermi level and decoupling of the 
 interband hopping of the two states,
 arising from the symmetry of the wave functions, plays 
 a crucial role in the effect.
\end{abstract}

\pacs{85.75.-d, 72.25.-b, 73.63.-b}

\maketitle

\section{INTRODUCTION}

The graphenes have been recently attracting much attention 
as a candidate material for spintronics devices because of 
its peculiar electronic 
structures,\cite{Ohishi,Tombros,Son,Kim,Li,Abanin,Yazyev} 
while so far most of promising materials in developing the 
devices have been found in ferromagnetic (FM) metals and FM
semiconductors such as GaMnAs.\cite{Zutic} 
In fact, recent experiments have achieved spin injection into 
graphene layers at room temperature for the first time among 
molecular materials,\cite{Ohishi,Tombros} 
and observed the magnetoresistance (MR) effect.\cite{Tombros}
In addition, several intriguing transport properties have been 
theoretically predicted especially for zigzag 
graphene nanoribbons (ZGNRs).\cite{Son,Kim,Li,Abanin,Yazyev,Karpan1,Karpan2,Ezawa,Martins,Guo} 
For instance, it is shown that ZGNRs
can be utilized even for generation of a spin polarized current 
in which an external electric field applied across ZGNR along 
the lateral direction may induce a half-metallic 
band structure being responsible for the spin polarized current.\cite{Son} 
Also an extraordinary large MR and generation of 
spin-polarized current are also theoretically found in a spin valve 
device consisting of ZGNR.\cite{Kim} The unique properties in the electronic 
transport properties of ZGNRs can be attributed to the characteristic 
band structures of ZGNR and the symmetry of wave functions near 
the Fermi level.\cite{Son,Kim,Li} 
Along this line, the family of graphenes related to ZGNR might be 
anticipated to exhibit unexpected electronic 
properties which can be useful in developing spintronics devices.
However, transport properties of ZGNRs under {\it finite} source-drain 
bias voltages $V_{\rm bias}$ have not been fully explored even 
from the theoretical points of view. 
For development of future device applications it is highly important 
to fully understand intrinsic behaviors in the transport properties 
of ZGNRs under a wide range of finite bias voltages.

In this paper we present a novel and intrinsic electronic transport 
property of ZGNR under finite bias voltage $V_{\rm bias}$,
based on first principle calculations, 
that a symmetric ZGNR with an antiferromagnetic (AFM) junction exhibits 
a dual spin filter effect under finite $V_{\rm bias}$, 
namely, the up- and down-spin electrons are unidirectionally filtered in 
the counter direction of $V_{\rm bias}$.
Based on an analysis using Wannier functions (WFs) 
and a tight-binding (TB) model, we further clarify that 
the spin filter effect arises from an interplay between
the band structure and absence of the interband hopping of
the $\pi$ and $\pi^*$ states near the Fermi level.

\section{COMPUTATIONAL DETAILS}

At each bias voltage $V_{\rm bias}$ the electronic structure 
of ZGNR shown in Fig.~1(a) 
is self-consistently determined under a temperature of 300 K by means 
of a non-equilibrium Green function (NEGF) method\cite{NEGF,Kondo} 
coupled with a local spin density approximation (LSDA)\cite{LSDA} 
in the density functional theory (DFT).\cite{KS} 
The equilibrium density matrix (DM) is evaluated by a contour integration 
method,\cite{Ozaki_FD} while the non-equilibrium DM is numerically 
computed on a line with an imaginary part of 0.01 eV 
which is parallel to real axis in complex plane.\cite{NEGF_Ozaki} 
The conductance and current are calculated by the Landauer 
formula.\cite{Landauer} 
Norm-conserving pseudopotentials are used in 
a separable form with multiple projectors to replace the deep 
core potential into a shallow potential.\cite{TM}
Pseudo-atomic orbitals (PAOs) centered on atomic sites are used 
as basis functions. The PAO basis functions, generated 
by a confinement scheme,\cite{Ozaki} are specified by H5.5-s2 and
C4.5-s2p2, where the abbreviation of basis functions, 
such as C4.5-s2p2, represents that C stands for the atomic symbol, 
4.5 the cutoff radius (Bohr) in the generation by the confinement 
scheme,\cite{Ozaki} and s2p2 means the employment of two primitive 
orbital for each of s- and p-orbitals. 
The real space grid techniques are used with 
the energy cutoff of 120 Ry as a required cutoff energy in numerical 
integrations and the solution of Poisson equation using FFT.\cite{SIESTA}
In addition, the projector expansion method is employed in the calculation
of three-center integrals for the deep neutral atom potentials.\cite{Ozaki2}
The geometrical structures used are optimized with a criterion of 
$10^{-4}$ hartree/bohr for forces on atoms under the periodic 
boundary condition. 
All the calculations were performed by an {\it ab initio} DFT code, 
OpenMX.\cite{openmx}

\section{RESULTS}

$n$-ZGNRs are characterized by the number of carbon atoms, 
even (symmetric) or odd (antisymmetric), in the sublattice being across 
ZGNR along the lateral direction, 
while Fig.~1(a) shows the case of eight carbon 
atoms abbreviated as 8-ZGNR. 
For ZGNRs we focus a spin configuration with an AFM junction 
as shown in Fig.~1(a). 
Considerable magnetic moments are found at both the zigzag edges
which can be attributed to the existence of the flat band 
near $X$-point.\cite{FM_Zigzag}
Although the AFM coupling between the zigzag edges is 
favored by about 10 meV per edge carbon atom compared to the 
FM coupling, we consider the spin configuration consisting of the FM
coupling between the zigzag edges and the AFM junction 
at the central region. 
The spin configuration is crucial for the spin filter effect we discuss, 
and might be realized by a magnetic field applied in a spin valve device
or chemical modifications for ZGNRs. 
It should be noted that the effect does not appear in the other spin 
configurations.
The current-voltage ($I$-$V$) characteristics for 8-ZGNR with the spin 
configuration depends on intriguingly not only spin, but also the 
direction of $V_{\rm bias}$ as shown in Figs.~1(b) and (c). 
Interestingly, the up-spin electron substantially flows only 
in the negative regime of $V_{\rm bias}$, while the down-spin electron
flows only in the positive regime.
The ratio of the spin dependent currents, $I_{\rm up}/I_{\rm down}$, 
turns out to be 44.3 at $-0.4$ V, and the value is equivalent to 
the rectification ratio for each spin-dependent current because 
of the $I$-$V$ characteristics.
The $I$-$V$ characteristics of other even cases, $n=$6 and 10, are also
found to be nearly equivalent. Thus, a symmetric ZGNR plays {\it dual} roles
as a unidirectional spin filter for {\it each} spin state under finite
$V_{\rm bias}$. 
The effect can also be regarded as a dual spin {\it diode} effect 
due to the unidirectional nature of the spin dependent current. 
In contrast, the current for 7-ZGNR with the spin configuration is 
almost independent of spin, and proportional to $V_{\rm bias}$ within 
the regime, leading to $I_{\rm up}/I_{\rm down}$ of a nearly one. 
Thus, it should be emphasized that the parity in the geometrical structure 
of $n$-ZGNR is the key factor even for the behavior of the spin 
polarized current as in the spin unpolarized current.\cite{Li}

The $I$-$V$ characteristics for 8-ZGNR is understood by examining
the dependency of the conductance on $V_{\rm bias}$. 
It is found in Fig.~2(a) that there is a conductance gap of about 0.4 eV
for both the up- and down-spin states around the chemical potential 
at $V_{\rm bias}=0$ V.
The gap for the up-spin state decreases as $V_{\rm bias}$ increases 
toward the negative direction, and approaches zero at about 
$V_{\rm bias}=-0.4$ V as shown in Fig.~2(b), while the zero gap
is kept up to $V_{\rm bias}=-0.55$ V. From then onwards the gap 
increases as illustrated in Fig.~2(c).
On the other hand, the gap for the down-spin state monotonically
increases as $V_{\rm bias}$ increases toward the negative direction. 
The conductance gaps for the up- and down-spin states are shown 
as a function of $V_{\rm bias}$ in the inset of Fig.~2(a).  
We see that the behavior of the gaps is opposite for the up- and 
down-spin states in the regime of the positive direction of
$V_{\rm bias}$ in contrast with that in the negative direction.
The shaded regions in Fig.~2(b) and (c) correspond to the regime in between 
two chemical potentials of the left and right leads, where the electron 
transmission can contribute to the current flow. 
Thus, it is confirmed that in the regime of the negative $V_{\rm bias}$,
$\vert I_{\rm up} \vert$ increases up to around $V_{\rm bias}=-0.55$ V,
where the gap opening occurs again, and from then onwards saturated.
Also it turns out that $\vert I_{\rm down} \vert$ should be nearly zero
in the regime of the negative $V_{\rm bias}$, since the shaded region
is always inlying in the conductance gap for the down-spin state.
In the regime of the positive $V_{\rm bias}$, the $I$-$V$ characteristics
can be explained in the same way by considering the reversed roles of
up- and down-spin states. Note that the reversal of the roles arises only in 
the spin configuration with the AFM junction.

The opening and closing of the conductance gap can be attributed to 
the band structure near the Fermi level of 8-ZGNR with the 
FM coupling between the zigzag edges. 
The exchange splitting $\Delta_{x}$ is found to be 0.553 eV 
at $X$-point from Fig.~3(a). The right blue and pink shades 
denote the conductance gaps for the up- and down-spin 
states in Fig.~2. It is clearly seen that the conductance gap
corresponds to an energy regime where the $\pi (\pi^*)$ state in the 
left panel overlaps with {\it only} the $\pi^* (\pi)$ state with 
the same spin in the corresponding right panel. 
The $\pi^*$ up-spin state in Fig.~3(a) overlaps with {\it only} the $\pi$ 
up-spin state in Fig.~3(d) in the conductance gap.
Once the overlap regime fades away at the situation given by
Figs.~3(b) and (e), and then onward it turns out that the $\pi$ up-spin
state in Fig.~3(c) overlaps with {\it only} the $\pi^*$ up-spin state in Fig.~3(f) 
in the regime of the conductance gap. 
On the other hand, for the down-spin state the energy regime, where 
the $\pi$ state in the left panel overlaps with {\it only} the 
$\pi^*$ state in the right panel, linearly increases as the energy
shift increases toward the negative direction as shown in
Figs.~3(d), (e), and (f). 
The same idea except for reversing of the roles of the $\pi$ and 
$\pi^*$ states can apply to the case of the energy shift toward 
the positive direction.
Since the band structures in the left and right panels can be regarded
as those of the left and right leads in Fig.~1(a), the correspondence
implies that the electron transmission from the $\pi (\pi^*)$ to 
$\pi^* (\pi)$ states is forbidden. In fact, it is shown that 
the Bloch function of the $\pi$ ($\pi^*$) state is antisymmetric (symmetric)
with respect to the $\sigma$ mirror plane which is the mid plane
between two edges.\cite{Li,Kim} 
Thus, the electron transmission
should be forbidden in the energy regime where the $\pi$ ($\pi^*$) 
overlaps with {\it only} the $\pi^*$ ($\pi$) states. 
The band structure of 7-ZGNR with the FM coupling between 
the zigzag edges is very similar to that of 8-ZGNR so that the above 
analysis can also apply to the case. However, as discussed later 
the transmission is allowed even for the energy regime, where {\it only}
the overlap between the $\pi$ ($\pi^*$) and the $\pi^*$ ($\pi$) 
states survives,
due to the absence of the mirror plane in 7-ZGNR, leading to 
the linear $I$-$V$ characteristics as shown in Figs.~1(b) and (c).

To further verify the physical origin, explained above, of the peculiar
$I$-$V$ curves which can be determined by the characteristic band
structure of ZGNRs and the symmetry of wave functions, we construct
WFs for the $\pi$ and $\pi^*$ states for 7- and 8-ZGNRs
with the non-magnetic state,\cite{WF} 
and evaluate tight-binding (TB) parameters using WFs.
It is confirmed that WFs of 7-ZGNR is neither 
symmetric nor antisymmetric (not shown), and that 
the absolute value of the nearest neighbor hopping integral 
between the $\pi$ and $\pi^*$ states is 0.485 eV which is comparable to 
the nearest neighbor hopping integral for the $\pi$ ($\pi^*$) state
of $-0.777$ (0.784) eV.
On the other hand, WFs for the $\pi$ and $\pi^*$ states
of 8-ZGNR are antisymmetric and symmetric with respect to 
the $\sigma$ mirror plane, and localized in three unit cells
along the ribbon direction as shown in Figs.~4(a) and (b). 
We confirm that the hopping integrals between WFs
for the $\pi$ and $\pi^*$ states are nearly zero (not shown). 
Also it turns out that the next nearest neighbor hopping integral and more, 
$h_n(n=2-4)$, are one order smaller than $h_1$ as shown in Table I. 
Thus, we can construct a simple TB model determined by a fitting,
containing the essence of the spin filter effect, which consists of 
only the on-site energy, the nearest neighbor hopping integral, 
and the exchange splitting $\Delta_{x}$. 
The fitted parameters in Table I are determined so that the band width 
and the degeneracy of the $\pi$ and $\pi^*$ states at $X$-point 
can be reproduced. It should be noted that the two band TB model
can be decomposed into two decoupled one band models due to the 
decoupling of the $\pi$ and $\pi^*$ states. The fact allows us to 
evaluate a spin polarized current for the $\pi (\pi^*)$ state 
using the Landauer formula\cite{Landauer} for 
the one band TB model and the fitted parameters as: 
\begin{eqnarray}
  I & =& \int dE (f_L-f_R) T(E),
\end{eqnarray}
where $f_L$ and $f_R$ are the Fermi functions with the chemical 
potentials for the left and right leads, respectively.
The transmission $T$ is given by an analytic formula:
\begin{eqnarray}
  T(E) &= &
   \frac{4S_{L}(E)S_{R}(E)}
        {\left[S_{L}(E)+S_{R}(E)\right]^2}
\end{eqnarray}
with
\begin{eqnarray}
  S_{L}(E) &=& \sqrt{4h_1^2
                -\left[E-(\varepsilon
                -\frac{1}{2}\Delta_{x})\right]^2},\\
  S_{R}(E) &=& \sqrt{4h_1^2
                -\left[E-(\varepsilon
                +\frac{1}{2}\Delta_{x}+V_{\rm bias})\right]^2},
\end{eqnarray}
where $S_{L}$ and $S_{R}$ are not zero only if the number 
in the square root is larger than zero, and therefore the transmission 
is finite within the energy range where both $S_{L}$ and $S_{R}$
survive. It is emphasized that the behaviors of the conductance
gap in Fig.~(2) can be easily traced by the model. 
Using Eq.~(1) we evaluate the gate voltage dependency of the 
$I$-$V$ characteristics for the up-spin state as shown in Fig.~4(c),
where $\Delta_{x}$ of 0.553 eV, being that of the spin polarized
8-ZGNR, is used, and the gate voltage $V_{\rm gate}$ is taken
into account by adding $V_{\rm gate}$ to the on-site energy.
The proposed TB model can be validated by the fact that 
the $I$-$V$ characteristics at $V_{\rm gate}=0.0$ V is quite similar 
to that in Fig.~1(b) calculated by the NEGF method.
In addition, it is found that the $I$-$V$ characteristics does not 
largely change as long as $V_{\rm gate}$ lies 
in between $\pm\frac{1}{2}\Delta_{x}$, while the absolute threshold 
voltage at which the current starts to flow can be tuned.
However, once $\vert V_{\rm gate} \vert$ exceeds 
$\frac{1}{2}\Delta_{x}$, the originally suppressed
current starts to leak in the positive $V_{\rm bias}$ regime.
The leak current is due to the fact that one of the chemical potentials
is located at the outside of the conductance gap. 
In case that ZGNR contacts with a metallic substrate, the Fermi 
level of ZGNR might be shifted by charge transfer between them. 
For such a case the gate voltage dependency apparently suggests that 
the Fermi levels must be adjusted within the energy 
regime of the exchange splitting by applying
a proper gate voltage in order to keep the high $I_{\rm up}/I_{\rm down}$. 

\section{CONCLUSIONS}

In summary, we demonstrate based on the NEGF method coupled with DFT
that the symmetric ZGNRs with an AFM
junction possess intrinsically a peculiar $I$-$V$ characteristics 
which can be regarded as a dual spin filter effect.
It is shown by analyzing the band structure, WFs of the $\pi$ 
and $\pi^*$ states, and a TB model that the physical origin 
of the spin filter effect can be attributed to the spin 
polarized band structure of the symmetric ZGNR and the symmetry of 
wave functions of the $\pi$ and $\pi^*$ states near the Fermi level. 
The dual spin filter effect might initiate a novel avenue 
in developing spintronics using graphene based devices.

\acknowledgments

This work is partly supported by CREST-JST, the Next 
Generation Supercomputing Project, Nanoscience Program,
and NEDO (as part of the Nanoelectronics project).

\newpage

\newpage
\noindent
{\bf Table:}\\

   \begin{table}[h]
     \caption{
        Tight-binding parameters (eV) evaluated by WFs
        denoted by $WF$, and a fitting, denoted by $fitted$,
        for the $\pi$ and $\pi^*$ states of the non-spin polarized
        8-ZGNR, where $\varepsilon$ is the on-site energy, and $h_1$, 
        $h_2\cdots$ the nearest and the second nearest neighbor 
        hopping integrals, and so on. The Fermi level is set to zero. 
       }
   \vspace{1mm}
   \begin{tabular}{lccccc}
   \hline\hline
               & $\varepsilon$ & $h_1$ & $h_2$ & $h_3$ & $h_4$ \\
     \hline
      $\pi$~~ (WF)     & -1.3609 & -0.7660 &  0.0076 &  0.0529 & -0.0352 \\
      $\pi^*$ (WF)     &  1.4486 &  0.7708 & -0.0400 & -0.0513 &  0.0269 \\
      $\pi$~~ (fitted) & -1.4165 & -0.7083 &  0      & 0       &  0 \\
      $\pi^*$ (fitted) &  1.4135 &  0.7067 &  0      & 0       &  0 \\
   \hline
   \end{tabular}
  \end{table}

\newpage
\noindent
{\bf Figure captions:}\\

Fig.~1\\
     (a) 8-ZGNR, terminated by hydrogen atoms, with an AFM junction 
     together with the spatial distribution of the spin density
     at $V_{\rm bias}=0$ V, where the red and blue colors stands for 
     positive and negative signs, and {\it L}, {\it R}, and {\it C}
     means the {\it left} and {\it right} leads, and the {\it central} 
     scattering region.
     $I$-$V_{\rm bias}$ curves for (b) the up-spin and (c) the down-spin
     states of 7- and 8-ZGNRs with the AFM junction.\\

Fig.~2\\
      Conductance of 8-ZGNR with the AFM junction with 
      $V_{\rm bias}$ of (a) 0.0, (b) $-0.4$, 
      and (c) $-1.0$ V. The inset
      in (a) shows the conductance gap near the chemical potentials.\\

Fig.~3\\
     (a)-(c) Band structures of 8-ZGNR with 
     a FM coupling corresponding to 
     the $left$ lead region in Fig.~1(a). 
     Band structures shifted with (d) 0.0, (e) $-0.4$, 
     and (f) $-1.0$ eV of 8-ZGNR with 
     a FM coupling corresponding to 
     the $right$ lead region in Fig.~1(a). 
     The blue and red shades show the conductance gap for the up- 
     and down-spin electrons, respectively. The horizontal dot line
     indicates the chemical potential, being position dependent, 
     of 8-ZGNR shown in Fig.~1(a).\\

Fig.~4\\
      Wannier functions for (a) the $\pi$ and (b) $\pi^*$ states 
      of the non-spin polarized 8-ZGNR.
      (c) $I$-$V_{\rm bias}$ curves of the up-spin state
      calculated by the simple TB model
      with the AFM junction for a series of the gate
      voltage $V_{\rm gate}$.\\










\begin{thebibliography}{99}
\bibitem{NEGF_Ozaki}
The details on the implementation of the NEGF method will be 
discussed elsewhere.
\end{thebibliography}
\end{document}